\documentclass[12pt]{article}

\usepackage{graphicx}
\usepackage{epsfig} 

\usepackage{scicite}

\usepackage{times}

\newenvironment{sciabstract}{%
\begin{quote} \bf}
{\end{quote}}

\newcounter{lastnote}

\title{Biological measurement beyond the quantum limit}

\author
{Michael~A.~Taylor$^{1,2}$, Jiri~Janousek$^3$, Vincent~Daria$^3$, Joachim~Knittel$^1$,\\ Boris~Hage$^3$, Hans-A.~Bachor$^3$, Warwick~P.~Bowen$^{2\ast}$\\
\\
\normalsize{$^1$Department of Physics, University of Queensland, St Lucia,}\\
\normalsize{ Queensland 4072, Australia}\\
\normalsize{$^2$Centre for Engineered Quantum Systems, University of Queensland,}\\
\normalsize{ St Lucia, Queensland 4072, Australia}\\
\normalsize{$^3$Department of Quantum Science, Australian National University,}\\
\normalsize{ Canberra, ACT 0200, Australia}\\
\\
\normalsize{$^\ast$To whom correspondence should be addressed. Email: wbowen@physics.uq.edu.au}
}

\date{}

\begin{document} 


\baselineskip24pt


\maketitle


\begin{sciabstract}
Quantum noise places a fundamental limit on the per photon sensitivity attainable in optical measurements. This limit is of particular importance in biological measurements, where the optical power must be constrained to avoid damage to the specimen. By using non-classically correlated light, we demonstrated that the quantum limit can be surpassed in biological measurements. Quantum enhanced microrheology was performed within yeast cells by tracking naturally occurring lipid granules with sensitivity 2.4 dB beyond the quantum noise limit. The viscoelastic properties of the cytoplasm could thereby be determined with a 64$\%$ improved measurement rate. This demonstration paves the way to apply quantum resources broadly in a biological context. 
\end{sciabstract}


Measurements of biological dynamics require low light levels to avoid photochemical intrusion upon biological processes and damaging effects on the specimen\cite{Peterman2003,Neuman1999}. With this constraint on optical power, the measurement sensitivity is fundamentally limited by noise due to the quantization of light\cite{Giovannetti2004,LIGO2011,Kolobov2000}. This {\it quantum noise limit}, commonly known as shot noise, can only be surpassed using quantum correlations between photons which increase the capacity of each photon to extract information\cite{Nagata2007,Treps2003,Brida2010}. Consequently, biology has long been viewed as an important frontier for quantum enhanced measurements\cite{Kolobov2000,Brida2010,Crespi2012}. However, this quantum limit has never been surpassed in a biological measurement; this includes recent quantum measurements such as quantum coherent intracellular tracking of NV nanodiamonds\cite{McGuinness2011}, where no quantum correlations were present between the detected photons. Two previous experiments have applied non-classically correlated photons to biological measurements without achieving quantum enhanced sensitivity; in these, optical coherence tomography was demonstrated within onion skin tissue\cite{Nasr2009} and protein concentrations measured\cite{Crespi2012}.

When measuring biological dynamics, one of the most versatile and powerful tools is laser based particle tracking. When used in conjunction with trapping in optical tweezers, this has allowed the manipulation of viruses and bacteria\cite{Ashkin1987}, unfolding of single RNA molecules\cite{Bustamante}, DNA sequencing\cite{Greenleaf}, and the discovery of step-like motion in the biological motor kinesin\cite{Svoboda1993} and muscle protein myosin\cite{Finer}. Another important and  particularly relevant application is real-time measurement of particle mobility within living cells\cite{Yamada2000}. Such measurements have revealed information about motor proteins, chemical gradients, and protein polymerization\cite{Senning2010}, and can even allow intracellular microrheology experiments which measure the viscoelasticity of the cytoplasm\cite{Norrelykke2004} and the viscoelastic changes during dynamic cellular proceses\cite{Selhuber2009}.

 Here we develop a new laser based microparticle tracking technique which is both immune to low frequency noise sources and specifically designed to operate with quantum correlated light. With this, quantum correlated light is used to perform microrheology experiments within {\it Saccharomyces cerevisiae} yeast cells,  verifying that the quantum noise limit can be overcome within living systems. The motion of naturally occurring lipid granules is tracked in real time and with sensitivity surpassing the quantum noise limit by 42$\%$ as they diffuse through the cytoplasm and interact with embedded polymer networks. This allows the viscoelastic moduli of the cytoplasm to be determined with a 64$\%$ higher measurement rate than possible classically. This laser tracking technique is widely applicable, extending the reach of quantum enhanced measurement to many dynamic biological processes. Furthermore, by demonstrating that biological measurements can be improved using quantum correlated light, our results pave the way to a broad range of applications in areas such as two-photon microscopy, super-resolution, and absorption imaging\cite{Giovannetti2004,Brida2010}.

\begin{figure}
 \begin{center}
   \includegraphics[width=8cm]{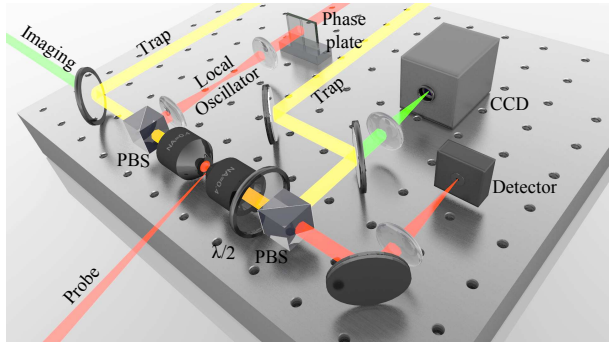}
   \caption{ Experimental layout.
   PBS: polarizing beamsplitter, $\lambda/2$: half waveplate.
   An Nd:YAG laser produces 10--500~mW of 1064~nm trapping field (yellow), which forms a counter-propagating optical trap to immobilize particles. Polarizing optics are used to isolate the trapping field from the detector. An imaging field (green) at 532~nm images the plane of the optical trap onto a CCD camera, allowing particles to be identified visually. A separate Nd:YAG laser produces the 1064~nm fields of the probe and local oscillator (red), which are used to measure particle position. The probe field, which illuminates trapped particles from the side, carries a strong amplitude modulation at 3.522~MHz for position measurement, and a weak phase modulation at 6.5~MHz which is used to generate an error signal for locking the phase between the probe and local oscillator. Probe photons which scatter from a trapped particle then interfere with the 100~$\mu$W local oscillator field. The local oscillator is shaped with a phase plate so that the interference between the scattered light and local oscillator maps the particle position to the transmitted light intensity which is detected on a bulk detector. 
}
 \label{Layout}  
 \end{center}
\end{figure}

In typical laser based particle tracking, the presence of a particle causes light to be scattered out of an incident field. The subsequent interference between scattered and transmitted fields manifests itself as a deflection of the incident field proportional to the displacement $x$ of the particle from the beam centre. This deflection is usually detected with a quadrant photodiode. The quantum noise limit is enforced by the probabilistic nature of photon detection events on either side of the photodiode. Quadrant photodetection is a special case of spatial homodyne detection\cite{Tay2009} where information contained in the field mode of interest is extracted via interference with a bright spatially shaped local oscillator field. In this framework, the quantum noise limit is given by $\Delta x_{\rm QNL} =  \eta^{-1/2}  n_{\rm scat}^{-1/2} \langle \psi_{\rm scat}' | \psi_{\rm det} \rangle^{-1}$,
 where $\eta$ is the detection efficiency; $n_{\rm scat}$ is the mean photon flux scattered from the particle which, for a centered Gaussian incident field, may be expressed in terms of the particles scattering cross-section $\sigma_{\rm scat}$, the incident photon flux $n_{\rm incident}$, and the incident beam width $w$ as $n_{\rm scat} = \sigma_{\rm scat} n_{\rm incident}/4\pi w^2$; $\psi_{\rm scat}$ and $\psi_{\rm det}$ are respectively the mode shapes of the scattered mode and a detection mode defined by the local oscillator field and the detection method; and $\psi_{\rm scat}' = \frac{d\psi_{\rm scat}}{dx} \big|_{x=0}$  in the limit of small particle displacement relevant to the work reported here (see supplementary information).

Assuming the correct phase is chosen between the local oscillator and the scattered field, the achievable sensitivity in optical tweezers based particle tracking experiments can be written in terms of the quantum noise limit as
\begin{equation}
\Delta x_{\rm meas} = \left[1 - \eta(1-V) \right]^{1/2} \times \Delta x_{\rm QNL} 
\end{equation}
where $V$ is the amplitude quadrature variance of the field within the detection mode $\psi_{\rm det}$ at the plane of the particle (see supplementary information). In the limit that the detected light is in a coherent state with $V=1$, the quantum noise limit is exactly reached. Using amplitude squeezed light, exhibiting non-classical photon anti-bunching, however, the variance $V$ may be suppressed below unity, allowing the quantum noise limit to be surpassed.

\begin{figure}
 \begin{center}
   \includegraphics[width=8cm]{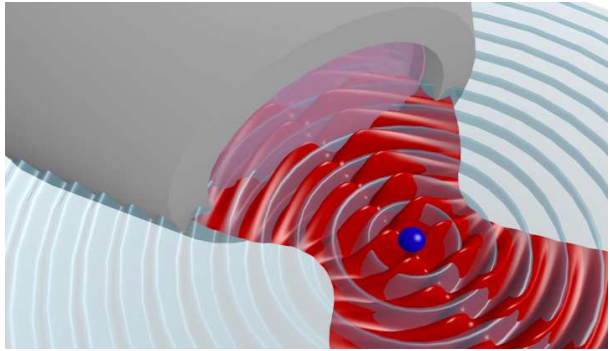}
   \caption{  Schematic of the particle tracking method.
    A trapped particle acts as the source of scattered light (faint blue). This scattered light is combined with the spatially antisymmetric local oscillator field (red), collected in an objective, and the interference is measured as intensity fluctuations. The phase of the scattered light is locked such that when the scattering particle is centered, the fields are $\pi/2$ out of phase. When the particle moves left, the scattered wavefront shift closer to the local oscillator field maxima on both the left and right, due to the spatial antisymmtry of the local oscillator. This leads to constructive interference; similarly moving right leads to destructive interference.  Hence the particle position is encoded on the detected light intensity.
}
 \label{LOmode} 
 \end{center}
\end{figure}

Two technical barriers have previously prevented squeezed light from being used in the context of biological measurements or particle tracking. First, such measurements are typically conducted at low frequencies, where classical noise sources constrain the possibility of generating squeezing\cite{McKenzie2004}; and second, after propagation through high numerical aperture lenses and biological samples, distortion prevents the spatial mode of the squeezed light from matching the detection mode. Here, we have developed a new modality of optical particle tracking to overcome these barriers, shown schematically in Fig.~1. Rather than relying on a single incident field to both interrogate the particle and to act as the local oscillator, two separate fields are used. A Gaussian {\it probe} field propagates transversely to the optical trapping axis, interrogating the particle and producing scattering, while a ``flipped" Gaussian {\it local oscillator} field, with a $\pi$ phase shift applied to one half of its transverse profile, propagates along the trap axis and acts to define the detection mode (see Fig.~2). Direct detection of the interference between the flipped local oscillator and scatterer light on a single photodiode provides equivalent particle position information to the quadrant photodiode in standard particle tracking. Now, however, the local oscillator field shape can be optimized independantly of the probe or trapping, and by directly amplitude squeezing the local oscillator field, any spatial mode perturbations occurring during optical propagation are applied equally to both the squeezing and local oscillator, ensuring perfect overlap at detection. Furthermore, the probe field can be stroboscopically pulsed without affecting the local oscillator. This allows a form of lock-in detection which shifts the particle position information to high frequencies, as has been proposed for squeezed light\cite{Yurke1987}. This is vital to our results, as squeezing is readily achievable at higher frequencies but notoriously difficult to attain in the range of many practical measurements\cite{McKenzie2004}.

 The local oscillator field was generated by an optical parametric amplifier, which when pumped at 532~nm, produced a 100~$\mu$W field with 6~dB of amplitude squeezing. A classical benchmark was produced by removing the pump and adjusting the optical power to match the 100~$\mu$W squeezed output, with the quantum noise limit reached at frequencies above 3~MHz.  To minimize degradation in the detected squeezing, all internal surfaces  were antireflection coated, with the apparatus measured to have a total of 19$\%$ optical loss. In order to characterize the stroboscopic measurement system, the probe illuminated a small defect in the sample chamber, producing scattered light to interfere with the local oscillator. The detector output was then studied with a spectrum analyzer, with traces shown in Fig.~3A for both squeezed and classical light. The amplitude modulation from the probe is visible as a peak at 3.522 MHz. At this frequency the quantum noise limit is achieved for classical light; while for squeezed light it is surpassed by 2.8 dB, corresponding to a detected squeezed variance of $V_{\rm det} = 1 - \eta (1-V) = 10^{-2.8/10} = 52\%$. At frequencies lower than 3 MHz, where typical optical particle tracking experiments operate, the noise floor is dominated by technical noise. Without stroboscopic measurement this noise would preclude reaching the quantum noise limit. It is worth noting that at low frequencies technical noise sources such as 1/f noise and laser noise are a common issue in conventional laser based particle tracking experiments\cite{Berg2004}. By removing this noise, the stroboscopic measurement technique demonstrated here may provide a path to improve these experiments even without utilizing non-classical light.

 \begin{figure}
 \begin{center}
   \includegraphics[width=8cm]{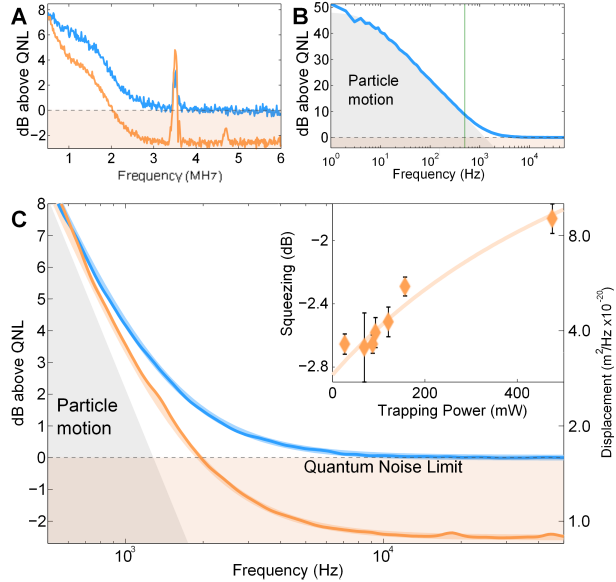}
   \caption{  Particle tracking spectra.
   Subplot {\bf a} shows the measured noise spectrum for classical (blue) and squeezed (orange) local oscillator without a trapped particle. The small peak visible at 4.7~MHz is caused by the modulations used for locking the laser. To measure mechanical motion, the detected signal was demodulated at 3.522~MHz and recorded with a sample rate of 100~kHz. Subplot {\bf b} shows a typical measured mechanical spectrum for a 2~$\mu$m silica bead. This closely follows the expected spectrum (light blue) for trapped Brownian motion. Squeezed light allowed the noise floor on this measurement to be lowered, with a typical measured spectra shown in {\bf c}. The inset shows the degradation in squeezing due to increasing trap power. This agrees well with a theoretical model with no fitting parameters, which assumes the small fraction ($7\times10^{-5}$) of trapping photons which reached the detector contributed shot noise.
 }
 \label{BeadData} 
 \end{center}
\end{figure}

 Initial particle tracking measurements were performed on the trapped thermal motion of 2~$\mu$m silica beads in water. Since the mechanical amplitude scales inversely with frequency, the motion above 1~kHz is difficult to resolve, as is typical of optical tweezers measurements. It is in this high frequency region that the simplistic model of Brownian motion can break down, and complex dynamic effects become significant\cite{Franosch2011}. This section of the spectrum is shown in Fig.~3c both with and without squeezed light. The squeezed light can clearly be seen to improve the sensitivity, and extend the frequency range over which the mechanical motion is resolvable, with a noise suppression of up to 2.7~dB, or 46$\%$.   As shown in the inset, the measured squeezing degraded as the trapping power increased as expected from theory.

 These results constitute the first demonstration of quantum enhanced particle tracking.  To demonstrate the scope of the technique, we performed microrheology experiments within {\em Saccharomyces cerevisiae}  yeast cells. It is known from intracellular measurements with a different yeast strain that the thermal motion of lipid granules is suppressed by networks of actin filaments within the cell cytoplasm, causing them to exhibit subdiffusive motion\cite{Yamada2000,Norrelykke2004}.    To study the granule motion in our experiments, the host cell was first immobilized by laser trapping with 170~mW of optical power, which also caused an estimated 1.5~K of cellular heating\cite{Peterman2003}. The shaped local oscillator was then used to extract a mechanical signal, which was particularly sensitive to small particles near the focus such as lipid granules. Larger structures, by contrast, produce a scattering profile with poor overlap with the local oscillator. Because of this, measurement of the scattered light extracted the motion of the lipid granules within the larger cell, rather than the bulk cellular motion. Similar to the bead tracking experiments, squeezed light improved the measured sensitivity such that it surpassed the quantum limit by up to 2.4~dB. If optical damage were a concern, this would allow the probe power to be reduced by 42$\%$.

 In measurements of diffusive motion, the key parameter of interest is generally the mean squared displacement (MSD). The MSD of a free particle undergoing thermal motion is
 \begin{equation}
 \Delta x^2(\tau) = \left< \left( x(t)-x(t-\tau) \right)^2 \right> = 2 D \tau^\alpha ,
\label{AlphaDef}
\end{equation}
 where $\tau$ is the delay between measurements, $D$ is the diffusion constant and $\alpha$ a diffusive parameter determined by the viscoelasticity of the surrounding medium. In a purely viscous medium, the particle exhibits Brownian motion, which is characterized by $\alpha=1$. The regime described by $\alpha < 1$ is subdiffusive motion, and is a signature of viscoelasticity in the surrounding medium with the ratio of loss to storage moduli of the medium given by\cite{Mason2000} $G''/G'=  {\rm tan} (\pi \alpha /2)$.  By measuring $\alpha$, the diffusive regime may be established, and information inferred about the local environment of the particle. The presence of a trapping potential or hard boundary causes the MSD to plateau at large delays\cite{Yamada2000}. The MSD was only analyzed here for delays below 0.01~s, where this effect was insignificant.

  \begin{figure*}
 \begin{center}
   \includegraphics[width=16cm]{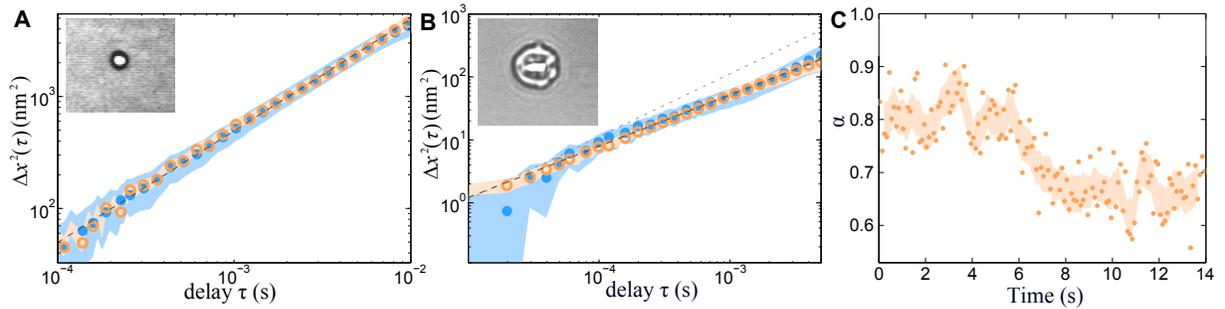}
   \caption{ Mean squared displacement data.
   Here we characterize the growth in displacement as a function of delay time. Typical results which were recorded for a duration of 0.1~s with silica beads and lipid granules in yeast are respectively shown in subplots {\bf a} and {\bf b}. Squeezed light measurements are shown as orange open circles, classical measurements as blue closed circles, and the shaded regions represent the measurement uncertainties. The dashed lines are linear fits to the data, which allow $\alpha$ to be determined. For the beads in water, this gives $\alpha=0.999\pm 0.006$, wheras it gives $\alpha=0.815\pm 0.008$ for the yeast results. In both cases this linear trend is followed for three orders of magnitude in delay, with a plateau appearing at longer delays (not shown).  For yeast, the data is clearly subdiffusive; for comparison, a diffusive trend is plotted (dotted line) alongside this fit.   Over all the measured data, $\alpha$ spanned from $0.6$ to $1$ with a mean of $0.81\pm0.01$.  Subplot {\bf c} shows the variation of $\alpha$ with time on a subset of data, with the measured values varying over sub-second timescales. 
   }
 \label{MSDplot} 
 \end{center}
\end{figure*}

 The MSD was extracted over a range of delay times for both silica beads and yeast results, with typical traces shown in Fig.~4 A and B respectively. The results from silica beads in water match the well known profile of diffusive motion, with $\alpha=0.994\pm 0.006$. By contrast, the results extracted with yeast cells reveal clearly subdiffusive motion with a non-stationary value of $\alpha$ varying between 0.6 and 1 as the lipid particles interact with different parts of the local environment (Fig.~4C), similar to other measurements in the literature~\cite{Selhuber2009}.   The lower noise floor from squeezing translates into a consistent reduction of measurement uncertainty. This is particularly visible in Fig.~4B at delays smaller than 40~$\mu$s, where the measured displacement is smaller than the classical measurement uncertainty. Squeezing was found to allow the diffusive parameter $\alpha$ to be determined with $22\%$ enhanced precision. Equivalently, this allows a 64$\%$ increase in the measurement rate while maintaining the same precision. Thus dynamic changes in $\alpha$ could be observed over shorter timescales, providing more information about the inhomogeneity of the local environment around the granule.

In the preceding results, the quantum noise limit was defined as the quantum limit on attainable sensitivity given the number of detected photons. Generally, not all of the scattered photons are detected, with some lost through optical inefficiencies. A more stringent definition is arrived at if 100$\%$ efficiency is allowed in the classical case, with $\Delta x_{{\rm QNL},~\eta=1} =  n_{\rm scat}^{-1/2} \langle \psi_{\rm scat}' | \psi_{\rm det} \rangle^{-1}=\eta^{1/2} \Delta x_{\rm QNL}$. Using this definition, and our overall efficiency $\eta=0.85$, the quantum limit is still surpassed by 1.7~dB when tracking lipid granules within yeast cells.

The results reported here demonstrate that squeezed light may be used to surpass the quantum limit on particle tracking sensitivity {\em per photon}. The absolute sensitivity achieved here was comparable to previous classical microrheology experiments, using a probe intensity of only 1$\mu$W/$\mu$m$^2$. The sensitivity could be directly improved by increasing the probe power and focusing it through the objectives\cite{Tay2009,Taylor2011}. In the limit of very high probe power, our technique should allow classical sensitivity approaching that achieved in recent non-biological experiments\cite{Chavez2008} of $10^{-28}$ m$^2$/Hz.  If optical losses were reduced to 10$\%$ and 10~dB of incident squeezing was used, as has recently been demonstrated in a number of experiments\cite{LIGO2011}, the quantum enhanced sensitivity would reach $2 \times 10^{-29}$ m$^2$/Hz.

The improved sensitivity available through squeezing could be useful to a wide range of applications. In microrheology experiments, the viscoelastic response of the medium may be probed on smaller timescales as the sensitivity improves, revealing both the properties of the cytoplasm and biological processes at higher frequency\cite{Buchanan2005}. The relevance of quantum enhancement in biology also extends beyond our improved mechanical sensitivity, as it can also be applied when generating images.
This holds promise in areas such as two-photon microscopy, super-resolution, and absorption imaging\cite{Giovannetti2004,Brida2010}.
%
%
 Several recent experiments have investigated the non-Brownian thermal motion of particles in water on very short time-scales, observing hydrodynamic memory\cite{Franosch2011,Lukic2005}, and the ballistic motion between collisions\cite{Huang2011}. However, the  instantaneous ballistic motion of a single particle in water remains undetected, as do elastic properties of fluid over very short timescales\cite{Huang2011}. Squeezed light would allow resolution of smaller displacements, which is vital in working towards such measurements. Another potential application for this technique is in optomechanical experiments with trapped levitating particles. Recent proposals predict that it may be possible to control the quantum mechanical state of such levitating particles\cite{Barker2010,Chang2010}. We have demonstrated that squeezing can be focused robustly in an optical trap. Introducing squeezed light to levitating optomechanical systems could enhance the measurement sensitivity, and even allows sensitivity beyond the standard quantum limit\cite{Walls}, which cannot be surpassed by increasing the light intensity. When strongly coupled to the mechanical motion, it may also enable the generation of non-classical states of motion of the trapped particle\cite{Jahne2009}.

{\bf Acknowledgements:} We would like to thank Ping Koy Lam for facilitating the squeezing experiments, Bill Williams and Nicolas Treps for useful discussions about microrheology and spatial squeezing respectively, and Magnus Hsu for input on the experiments. This work was supported by the Australian Research Council Discovery Project Contract No. DP0985078.

\end{document}